\pdfoutput=1

\documentclass[10pt,twocolumn,conference]{IEEEtran}

\usepackage{caption, amssymb, amsmath, graphicx, algorithm, algorithmic}
\usepackage[numbers,square]{natbib}
\usepackage{url, caption, subfigure, color, comment, multirow}
\usepackage{parcolumns, balance, multicol, listings}
\usepackage[autostyle,english=american]{csquotes}
\usepackage{amsmath}
\usepackage{amssymb}
\MakeOuterQuote{"}

\begin{document}

\title{Development of a Burst Buffer System for Data-Intensive Applications}

\author{
\begin{tabular}{ccccc}
Teng Wang$^\dag$ & 
Sarp Oral$^\ddag$ & 
Michael Pritchard$^\dag$ & 
Kevin Vasko$^\dag$ & 
Weikuan Yu$^\dag$ 
\end{tabular} 
\hspace{1pc}~\\
\begin{tabular}{cc}
    Auburn University$^\dag$  & Oak Ridge National Laboratory$^{\ddag}$ \\
      Auburn, AL 36849 & Oak Ridge, TN 37831 \\
    {\{tzw0019,mjp0009,vaskokj,wkyu\}@auburn.edu} & {\{oralhs\}@ornl.gov} \\  
\end{tabular}
}

\maketitle
\thispagestyle{empty}

\begin{abstract}
Modern parallel filesystems such as Lustre are designed to provide high, scalable I/O bandwidth in response to growing I/O requirements; however, the bursty I/O characteristics of many data-intensive scientific applications make it difficult for back-end parallel filesystems to efficiently handle I/O requests. A burst buffer system, through which data can be temporarily buffered via high-performance storage mediums, allows for gradual flushing of data to back-end filesystems. In this paper, we explore issues surrounding the development of a burst buffer system for data-intensive scientific applications. Our initial results demonstrate that utilizing a burst buffer system on top of the Lustre filesystem shows promise for dealing with the intense I/O traffic generated by application checkpointing. 
\end{abstract}

\section{Introduction}
\label{sec:intro}
One consequence of the increasing node quantity on supercomputing systems is a corresponding rise in storage system bandwidth demand; unfortunately, the current growth rate of supplied I/O bandwidth does not achieve parity with increases in computational power and node count. For example, the 2012 upgrade of the Jaguar supercomputer housed at the Oak Ridge Leadership Computing Facility (OLCF) to the Titan supercomputer~\cite{titan} enhanced the system's computing power from 1.3 Pflops/s to 20 Pflops/s -- an improvement by a factor of over 15. The corresponding upgrade to OLCF's Spider filesystem~\cite{oralolcf} saw only a four-fold increase in aggregate bandwidth (240 Gb/s to 1 TB/s). In observing this trend of growth disparity, we can see that data-intensive applications are becoming increasingly I/O bound. 

An interesting characteristic of scientific applications is that their I/O behavior tends to follow a bursty pattern~\cite{wang2004file}. This bursty I/O arises from the employment of checkpointing by many scientific applications. We can further characterize the execution of scientific applications as a two-phase cycle consisting of computational periods followed by periods of I/O in which checkpoint data is offloaded to the underlying parallel filesystem (PFS). Since these two phases occur sequentially, one approach to optimize performance is minimizing the I/O phase. Our approach to achieve this optimization is utilization of a \emph{burst buffer system} to quickly handle the bandwidth-intensive I/O requests generated by checkpointing.

A burst buffer is a large buffering space provisioned by one or more levels of high-performance storage (e.g. DRAM, SSDs). By using burst buffers, scientific applications can quickly flush checkpoint data to a temporary buffer, thus removing expensive PFS write operations from the critical path of execution. This in turn allows the next phase of computation to execute in parallel with the gradual flushing of data from the buffer. Though realizing this parallelization is straightforward, there are additional design considerations that should be taken into account. We divide these considerations into client-side and server-side categories. From the client's side, we want to facilitate rapid crash recovery for applications. Additionally, the flushing of checkpoint data to the buffer (and eventually the PFS) should be handled in a balanced and coordinated manner to maximize performance gains. For the server, an effective fault tolerance framework is necessary to protect the application from node failures within the burst buffer system. 

In this paper, we propose the design of an efficient and reliable burst buffer system taking into account the aforementioned design considerations. Our initial experiments on top of the Lustre filesystem reveal it is able to improve I/O performance by a factor of 2.78.

\section{A High-Level Overview of Burst Buffer System}
\label{sec:design}

\begin{figure}[htb]
  \begin{center}
  \vspace{0.0pc}
  \includegraphics[width=0.95\columnwidth]{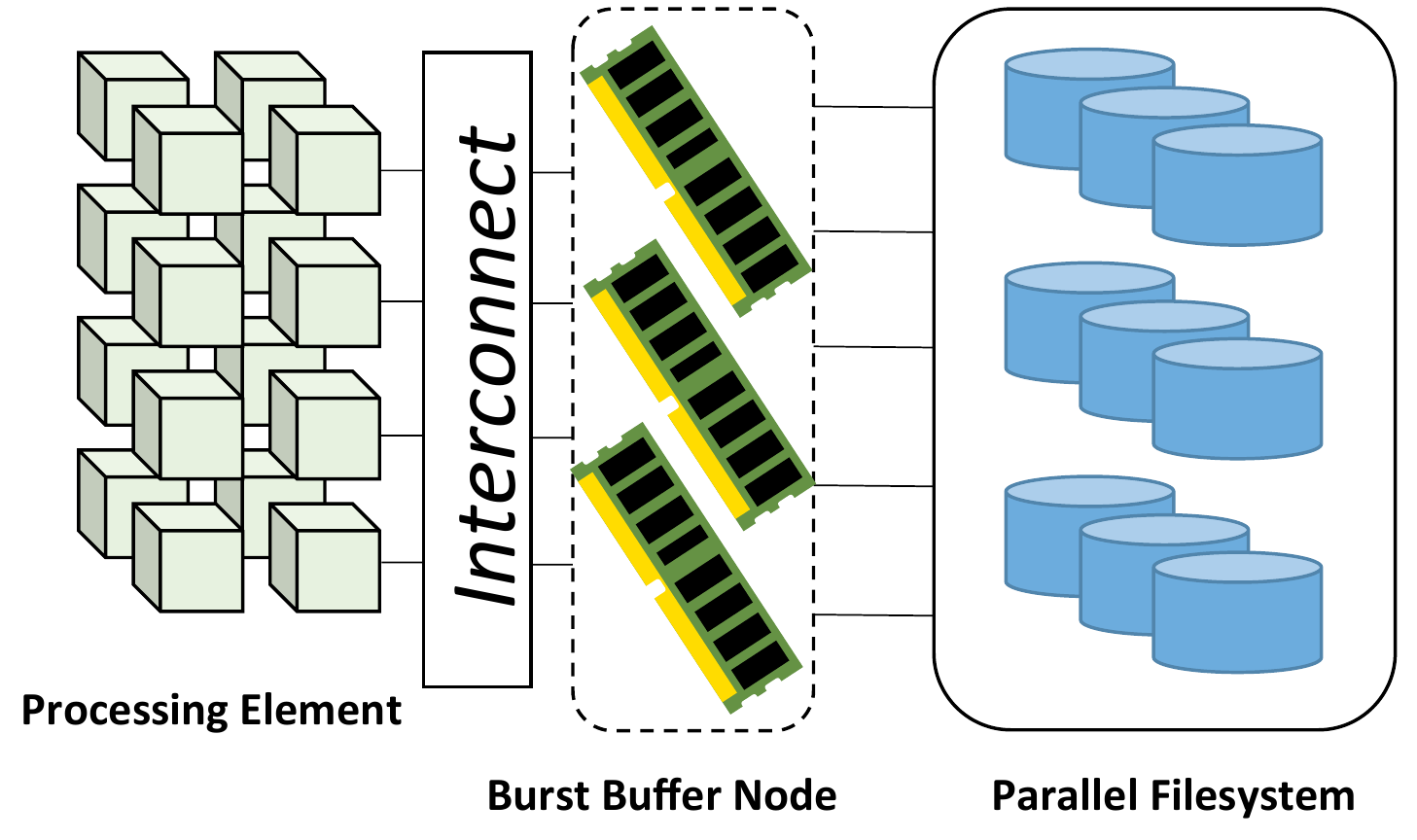}
  \caption{Architecture of a burst buffer system}
  \vspace{-1.0pc}
  \label{fig:burstmem}
  \end{center}
\end{figure}

As seen in Fig.~\ref{fig:burstmem}, burst buffer nodes typically lie between the processing elements and back-end PFS. A high-speed interconnect links the processing elements and burst buffer nodes in order to maximize data throughput.

Our burst buffer system consists of three entity types: client, server, and manager. Clients reside on each compute node and provide APIs that expose a simple key-value store interface to parallel jobs. Servers exist as a daemon on each burst buffer node and are logically linked using a ring topology. These daemons store the key-value pairs generated by the client and handle data flushing to the PFS. Finally, the manager is a singular entity collocated on one of the burst buffer nodes. It handles the initialization and maintenance of the server ring.

The inclusion of a burst buffer system allows scientific applications to
follow
a new checkpointing flow. Following each phase of computation, the
application
coordinates the transfer of checkpointing dataset to the servers using the
client-side API. After client-side API selects
the server for each key-value pair based on consistent
hashing~\cite{karger1997consistent}, it transfer all the 
key-value pairs to servers. Following this transaction, the 
servers assume the task of flushing the dataset to the PFS, freeing 
the compute nodes to begin the next cycle of computation. 

\section{Client-Side Performance Considerations}
In considering client-side performance, we examine three factors: balanced offloading of data, coordinated data flushing from server nodes to the PFS, and rapid application crash recovery. This section discusses how the burst buffer system addresses each of these factors.

\subsection{Coordination for Balanced Data Buffering Service}
Though a burst buffer system can accelerate scientific applications' I/O bandwidth by buffering checkpointing datasets in DRAM and SSDs, unbalanced workloads can significantly limit this benefit. For severely unbalanced workloads, some
burst buffer servers may experience significant data spillover to secondary storage (e.g. SSDs). Performance in these cases is limited by the servers with heavy workloads. Coordinated load balancing is introduced to minimize workload imbalances, thus maintaining optimal performance. To achieve this balance, overloaded servers redirect clients to underutilized servers. Servers are considered overloaded when they no longer have sufficient memory (e.g. DRAM) to buffer a write request.

When an overloaded server receives a write request, it begins issuing requests to neighboring servers querying their available memory. This message propagation is realized using the servers' logical ring topology. After receiving the responses, the overloaded server identifies the server with the lowest memory utilization and forwards this information to the client. This approach incurs some inter-server communication overhead, but offers more consistent ingress bandwidth.

\subsection{Coordination for Data Flushing}
In most cases, each burst buffer server possesses noncontiguous segments belonging to the same shared files. Directly flushing this data to the PFS can result in significant lock contention. Our burst buffer system avoids this issue by utilizing two-phase I/O~\cite{thakur1999data}. In the communication phase, all processes exchange metadata for their buffered write requests and figure out the size of each shared file. Each shared file is then logically partitioned into $n$ file domains (where $n$ is the number of servers), each of which is assigned to a single server. All the servers shuffle their write requests so that each server acquires all the data within its own domain. In the I/O phase, all the servers flush the write requests to the PFS.

\subsection{Coordination for Application Restart}
A checkpoint file is not modified once written -- it is only read during a restart after a failure~\cite{rajachandrasekar20131}. Although this is assumed to be a rare event compared to the number of checkpoints taken, our burst buffer system is designed to support checkpoint dataset retrieval from the burst buffer without touching the PFS. This is achieved by preserving recent checkpoint datasets so that applications can easily
rollback to a previous state. 

After data shuffling, each server maintains a lookup table that records the filename and global size of each buffered file. These two parameters determine the distribution of file segments among servers according to the file domain established during data shuffling. Given a retrieval request, each server is able to tell which server hosts the requested data. 
With this lookup table, each server is able to locate the peers necessary to service read requests from clients. Once the client acquires this information, it retrieves its requested dataset from the host servers.

\section{Server-Side Data Resilience}
\label{server_fault}
Data resilience among burst buffer servers is facilitated through the use of a logical ring topology. This topology, which is initially established through the manager, is maintained through periodic synchronizations among burst buffer servers. In this section, we present our strategies for server synchronization and data replication.

\subsection{Synchronization for Failure and Membership Detection}
\label{server_sync}
Prior to any synchronization operations, burst buffer servers must initialize the ring topology through the manager. During initialization, each server sends an init message to the manager. After a set waiting period, the manager arranges the ring and distributes this information (including server IDs) to each server and client. 

The synchronization mechanism itself is based on the Chord protocol~\cite{stoica2001chord}. The Chord protocol distributes the task of maintaining consistency to each server in the ring. Since each server need only store part of the ring (i.e. its predecessor and successors), synchronization message size is reduced. Additionally, the overhead involved with broadcasts is avoided by having each server only synchronize with its neighbors.
 
\begin{figure}[htb]
\begin{center}
\vspace{-0.5pc}
\includegraphics[width=0.40\textwidth]{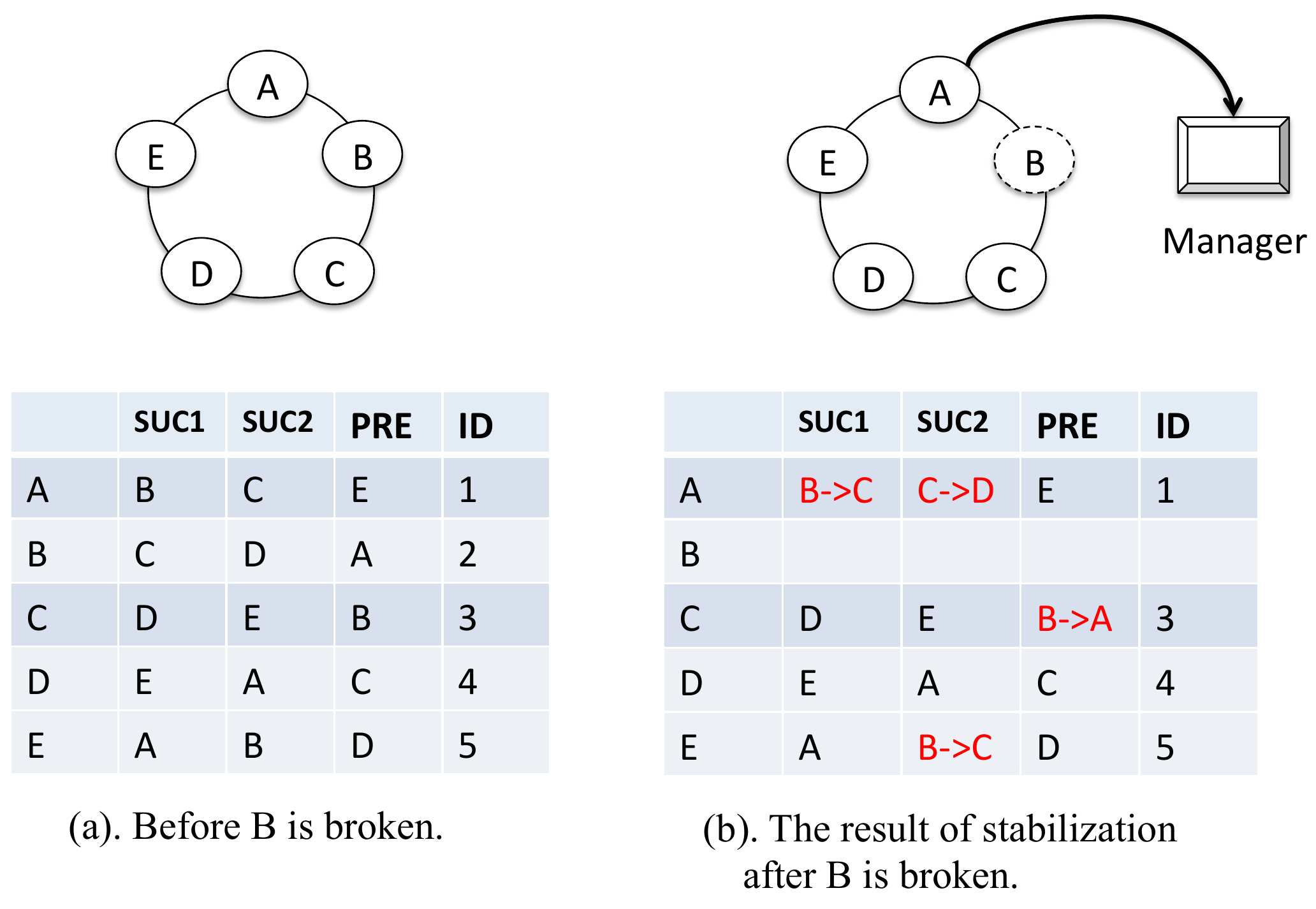}
\caption{Synchronization during Server Failure}
\vspace{-0.5pc}
\label{nodefailure}
\end{center}
\end{figure}

Two situations require modifications of the server ring: server failures and server joins. Periodic stabilization amongst neighboring servers handles these two cases. Fig.~\ref{nodefailure} demonstrates an example case where Server B fails. In this example, each server maintains a single predecessor node (PRE) as well as two successors (SUC1, SUC2). During stabilization, server A will attempt to contact server B. Since server B is down, the contact will fail. As a result, A will remove B from its successor list and contact C to inform it of B's failure. Additionally, server A will update its successor list by querying C for its successors. Corresponding updates will occur for each server whose neighbor lists are affected by B's failure. Finally, the manager is informed of B's failure.

\begin{figure}[htb]
\begin{center}
\vspace{-0.5pc}
\includegraphics[width=0.40\textwidth]{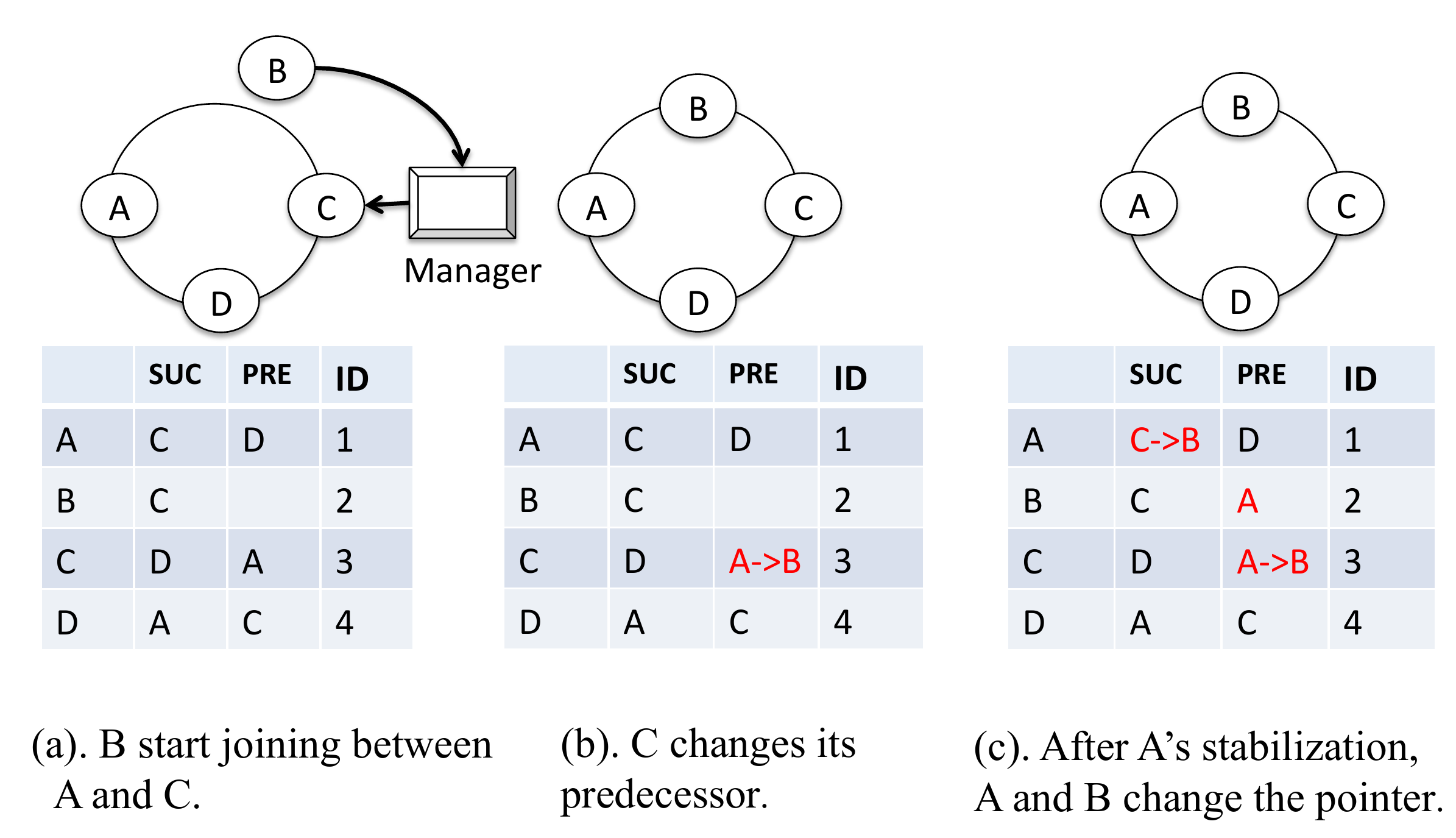}
\caption{Synchronization during Server Join}
\label{nodejoin}
\vspace{-0.5pc}
\end{center}
\end{figure}

Fig.~\ref{nodejoin} shows an example of a server joining the ring. In this example, each server is maintaining only a single successor and predecessor. Server B informs the manager of its intent to join the ring with server C as its predecessor. The manager then informs C of B's introduction to the ring. This triggers a stabilization operation in which, as with server failures, each affected server appropriately updates its neighbor lists.

In addition to maintaining the freshness of server neighbor lists, the stabilization procedure also handles updates to the data replication scheme mentioned in the next section.

\subsection{Fault Tolerance Data Management}
\subsubsection{Data Protection via Replication}
\begin{figure}[htb]
\begin{center}
\includegraphics[width=0.40\textwidth]{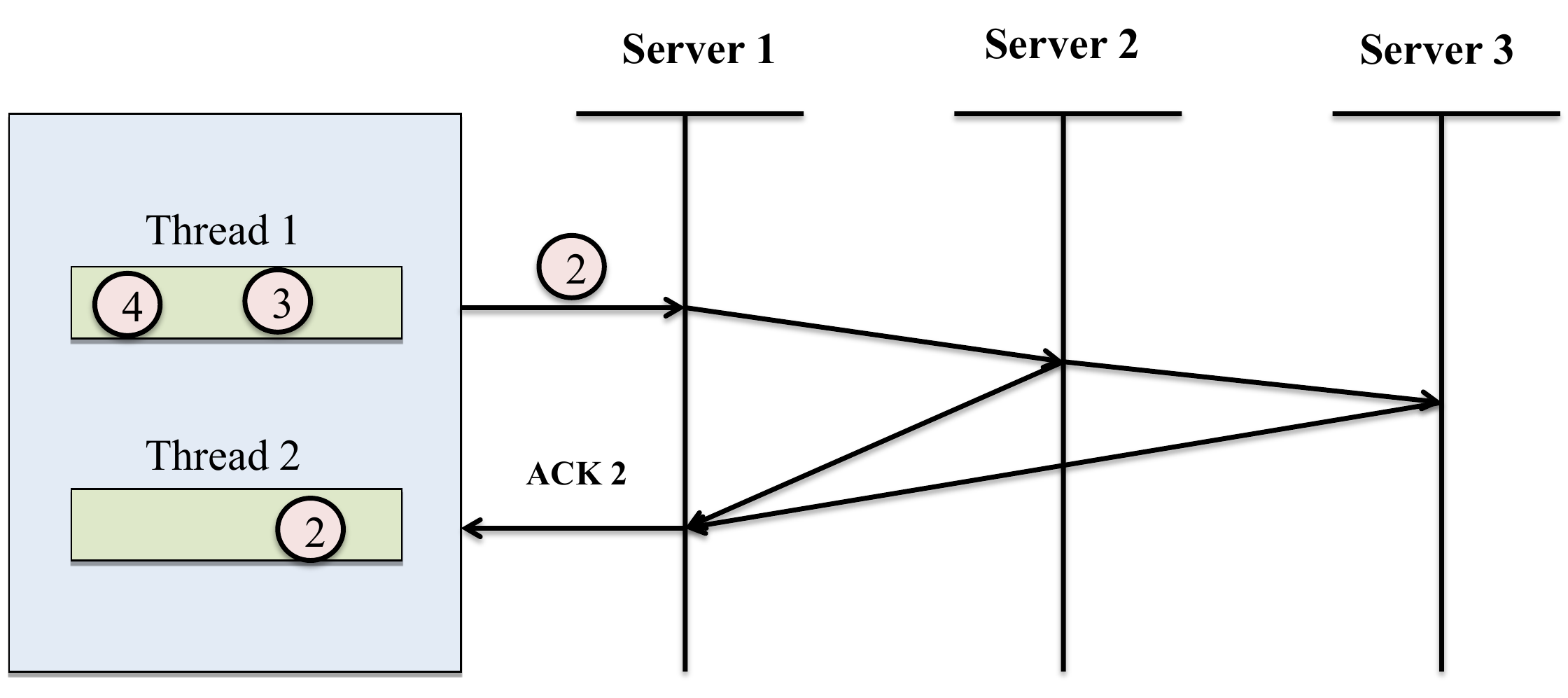}
\caption{Interaction during Replication}
\vspace{-0.5pc}
\label{replicating}
\end{center}
\end{figure}

Fault tolerance in the burst buffer system is achieved through data replication along successors in the ring. Fig.~\ref{replicating} provides an example of replication when two successors are maintained. When the data (key-value pair 2) arrives at server 1, the data is stored and forwarded to server 2. Similarly, server 2 stores and forwards the data to server 3. After receiving the data, each of the successors issues an ACK back to server 1. A final ACK is forwarded to the client informing it of the successful transfer of data. From the client's perspective, thread 2 manages ACKs. After the key-value pair is sent, it is buffered on thread 2's queue. The key-value pair is removed from the queue after server 1's ACK is received. By managing ACKs asynchronously, multiple key-value pairs can be issued concurrently, improving pipeline utilization.

\subsubsection{Data Recovery under Server Failure}
Server failure can be detected by either server-side stabilization or the client. Timeouts occurring after sending a key-value pair trigger a server failure detection by the client. Following the initial timeout, the client asks the targeted server's predecessor to confirm the failure. A confirmed failure will result in the manager being informed, after which an updated server list will be issued to the clients. The confirmation will also trigger the ring stabilization procedure described in section IV-A. 

\section{Initial Experiment Evaluation}
We implemented an initial prototype of the burst buffer system to evaluate its potential for I/O acceleration. We use log-structured writes to log all the write requests to the DRAM and SSD. The data transfer is implemented with 
CCI (Common Communicaiton Interface)~\cite{atchley2011common}, which is able to leverage the native transport protocols of diverse interconnects, such as Cray GNI for Cray Gemini and IB verbs for Infiniband. More details can be found in ~\cite{wang2014burstmem}. We have also implemented two ways of consistent hashing for data placement, and compared their performance. The first strategy assigns the key-value pairs using the Ketama algorithm~\cite{ketama}. With this approach, the key-value pairs generated by each client are distributed to all servers in a load balanced manner. The second strategy assigns each server a distinct set of clients, all traffic from which is directed to the server.

\subsection{Methodology}

{\bf Testbed}:
Our experiments are conducted on the Titan supercomputer. Each node is equipped with a 16-core 2.2GHZ AMD Opteron 6274 (Interlagos) processor, 32 GB of RAM, and a Cray custom high-speed interconnect. In all the experiments, we pinned 16 MB DRAM buffer for CCI data transfer among two communication entities. Since there are no I/O nodes deployed for I/O buffering, we use a separate set of compute nodes to act as burst buffer servers. Out of the 256 allocated to the experiment, 128 of the compute nodes are used as clients that write data into the burst buffer server. The other 128 compute nodes are allocated as the burst buffer servers. In every experiment, we place one process on each physical node. 

Titan is connected to Spider II, a center-wide Lustre-based file system. It offers 1 TB/s aggregated bandwidth. The default stripe size of each created file is 1 MB. The default stripe count is 4.  

{\bf Benchmarks}:
To evaluate the performance of burst buffer, we employed a synthetic workload using IOR~\cite{ior} and reported the average of 5 test results.

IOR is a flexible benchmarking tool that is able to emulate diverse I/O access patterns. We added burst buffer support to IOR by redirecting all writes from the processes to the burst buffer system instead of the PFS. This new version of 
IOR is referred to as BB-IOR. We refer to BB-IOR with the two data placement strategies respectively as BB-IOR-Ketama and BB-IOR-ISO.	To emulate bursty I/O behavior, we set {\em interTestDelay} to 20 seconds between any two I/O phases, and iterated 10 times.

\subsection{Ingress Bandwidth}
\label{sec:ingress-test}

\begin{figure}[htb]
  \begin{center}
  \vspace{-0.0pc}
  \includegraphics[width=0.36\textwidth]{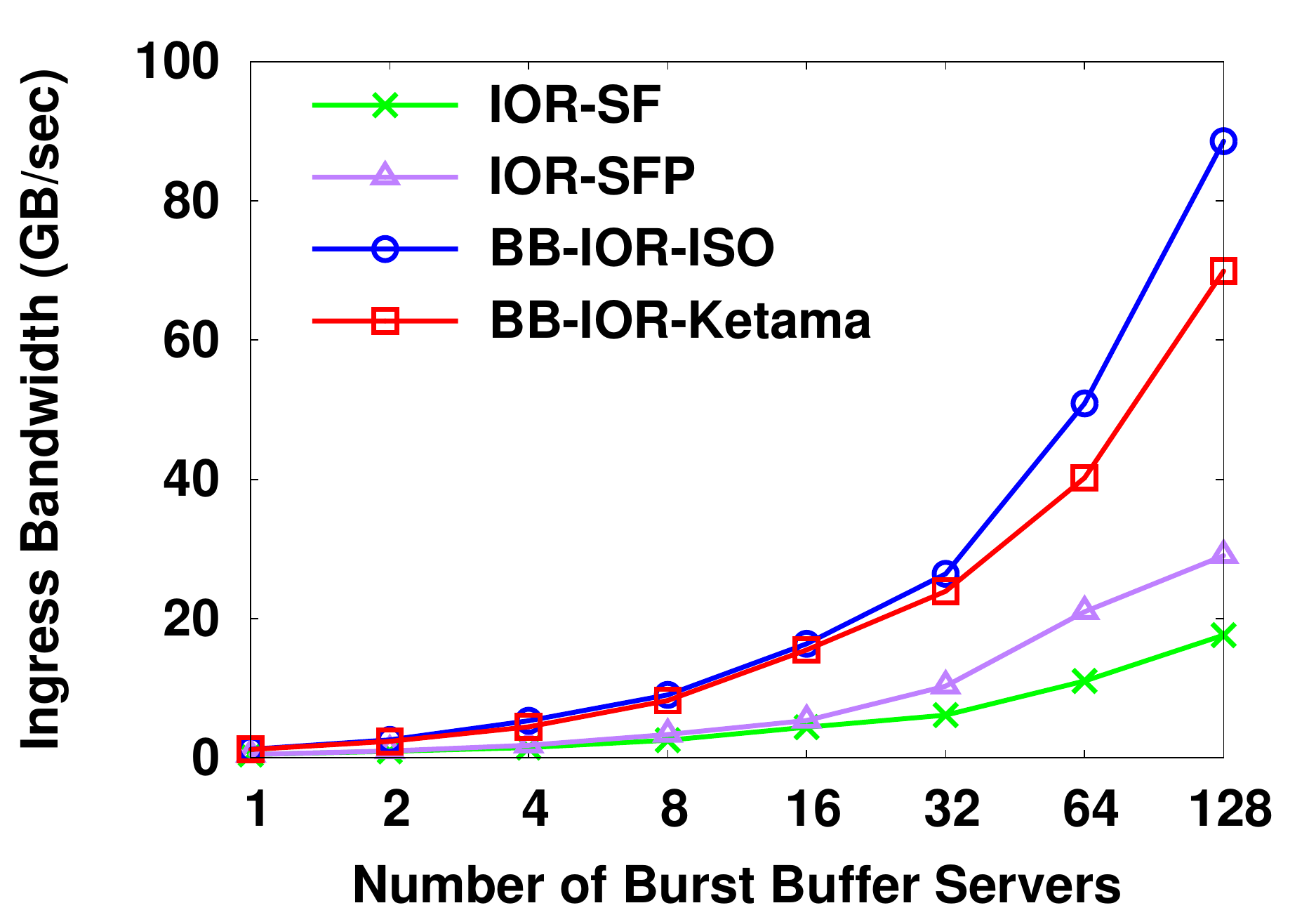}
  \caption{Ingress I/O Bandwidth with Increasing Number of Burst Buffer Servers}
  \vspace{-1.0pc}
  \label{fig:cmp}
  \end{center}
\end{figure}

We first investigated the ingress bandwidth that burst buffer can provide to absorb write requests. We used the IOR benchmark and increased the number of burst buffer servers from 1 to 128. In each test, we used an equal number of IOR clients and burst buffer servers to stress the system. We used a 1MB (default stripe size) transfer unit to alleviate lock contention issues in the Lustre filesystem. IOR-SFP and IOR-SF respectively represent two common write patterns: separated file per process and shared file. In the IOR-SF case, we set the stripe count to the number of clients. For comparison, we set the stripe count to 1 for IOR-SFP case, so the same number of Object Storage Targets (OST) are utilized as the number of clients. On average, each IOR client wrote 4 GB data.

Figure~\ref{fig:cmp} compares the ingress bandwidth IOR received with and without burst buffer support. Both BB-IOR cases yielded significant performance improvement over the original IOR. In particular, BB-IOR-ISO delivered the
highest bandwidth. It achieved 278.2\% and 174.5\% improvement on average, when compared to IOR-SF and IOR-SFP respectively. This improvement is consistent across different numbers of burst buffer servers. We also observed that the cumulative bandwidth of BB-IOR-ISO improved proportionally with the number of processes. In comparison, the bandwidth growth was much slower for BB-IOR-Ketama, IOR-SPF and IOR-SF. This is because BB-IOR-ISO localized each process's writes on one server; hence, it balanced the I/O traffic on each server. We believe the isolated data placement strategy is a good choice for burst buffer service.

\subsection{Evaluation of Burst Buffer with Hybrid Storage}
\begin{figure}[htb]
\centering
\vspace{-0.5pc}
\includegraphics[width=0.40\textwidth]{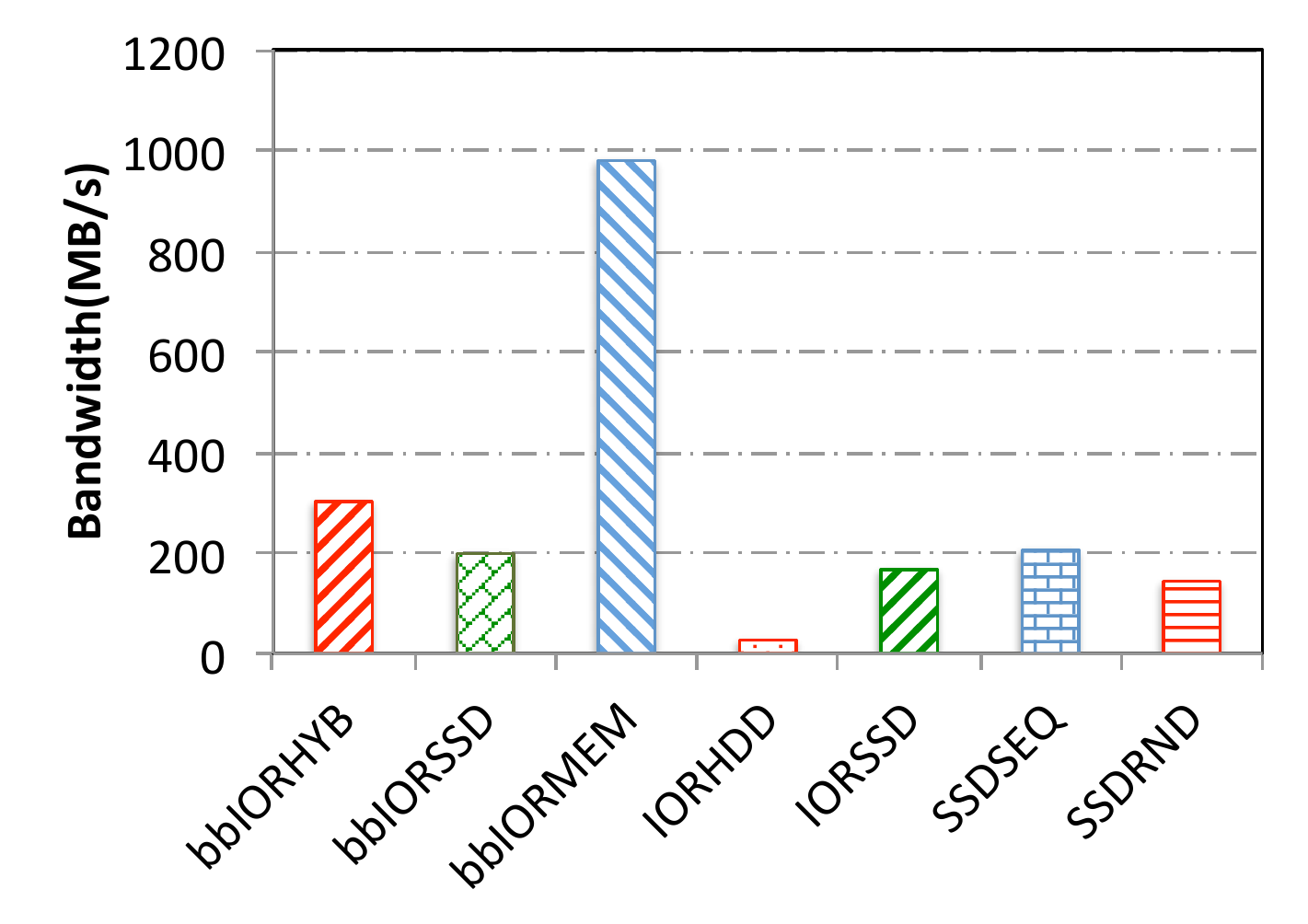}
\caption{Performance of HLW}
\vspace{-0.5pc}
\label{hybwrite}
\end{figure}

One important feature of burst buffer is its ability to leverage hierarchical storage devices. We evaluated this feature on our in-house cluster that contains 12 nodes equipped with dual-socket quad-core 2.13GHZ Intel Xeon processors, 8GB of DDR2 800MHZ memory, 7200 RPM HDD and Linux 2.6.18-358.18.1.el6. All nodes were interconnected by Mellonox QDR 4X Infiniband. Two nodes were equipped with OCZ-VERTEX4 SSDs yielding 500MB/s theoretical peak throughput. The SSDs were mounted with the EXT3 filesystem. For performance evaluation, we used two processes on one node, each dumping 2GB data to a burst buffer server on another node equipped with SSD. We used Infiniband verbs as the 
CCI transport. Figure~\ref{hybwrite} demonstrates the result.

In the first experiment (bbIORMEM), we allocated 4GB of memory on the burst buffer server so that data from both processes could fit in memory. In the second experiment (bbIORSSD), we didn't allocate any memory on the server, so
that all data spilled to the SSD following log-structured write. In the third experiment (bbIORHYB), we allocated 2GB memory on the server, and let the remaining 2GB data spill to the SSD. We measured the ingress bandwidth of the three cases by having each client write on a shared file with a transfer size of 16KB. For comparison, we measured the throughput of two processes dumping the same amount of data directly to a local SATA disk (shown as IORHDD) and
SSD (shown as IORSSD) without going to burst buffer. In addition, we listed the SSD's sequential and random write performance (respectively shown as SSDSeq and SSDRND) for single process write. For all cases, bbIORMEM yielded the best performance (980MB/s). The performance of bbIORHYB (302.29MB/s) was second to bbIORMEM since half the data reached SSD. We also observed that the performance of bbIORSSD (198.83MB/s) is comparable to SSDSEQ (205.99MB/s), which is reasonable for a log-structured write pattern. On the other hand, we observed much worse performance for IORHDD (27.11MB/s) and IORSSD (166.7MB/s). Lacking burst buffers, the two clients' writes were issued directly to the SSD. From the SSD's perspective, the arrival order of the write requests followed a semi-random order, which was less favored. IORHDD also suffered from the same issue, but its poor performance compared to IORSSD was because of the lower throughput of the SATA disk.

\section{Conclusion}
In this paper, we characterize major challenges in designing a burst buffer system, including the demand for timely application checkpointing/restart services and data resilience in the face of burst buffer server failure. Accordingly, we
propose the design of an efficient, reliable burst buffer system that aims to overcome these challenges. Our initial results demonstrate that our burst buffer prototype on top of the Lustre filesystem is very promising in absorbing intensive I/O traffic from application checkpoints. 

As future work, we plan to implement server-side orchestration for fault tolerance and provide load-balancing and application restart support for the client.

\balance
\small {
\bibliographystyle{plain}
\bibliography{paper}
}

\end{document}